# Discomfort: A New Material for Interaction Design

m.c. schraefel, University of Southampton

Michael Jones, Brigham Young University

ABSTRACT

This paper proposes discomfort as a new material for HCI researchers and designers to consider in any application that helps a person develop a new skill, practice or state. Discomfort is a fundamental precursor of adaptation and adaptation leads to new skill, practice or state. The way in which discomfort is perceived, and when it is experienced, is also often part of a rationale for rejecting or adopting a practice. Engaging effectively with discomfort may lead to increased personal development. We propose incorporating discomfort-as-material into our designs explicitly as a mechanism to make desired adaptations available to more of us, more effectively and more of the time. To explore this possibility, we offer an overview of the physiology and neurology of discomfort in adaptation and propose 3 issues related to incorporating discomfort into design: preparation for discomfort, need for recovery, and value of the practice. We look forward in the Workshop to exploring and developing ideas for specific Discomfortable Designs to insource discomfort as part of positive, resilient adaptation.

## 1 INTRODUCTION AND MOTIVATION

This paper proposes that discomfort can be used as a new deliberate and explicit design material in HCI. We make this proposal based on physiologic and neurologic consideration of the role of discomfort in adaptation. Adaptation is the process of physiological changes invoked via our metabolism to maintain homeostasis in the face of environmental demands. This adaptation is part of our innate plasticity [4]: our ability to respond to a stimulus, and to develop the functional capacity or state that the stimulus triggers. Adaptation affects every organ system in our bodies – from nerve cells in our brains to the bacteria in our guts.

A considerable challenge we encounter anytime we want or need to develop an adaptation is that the demands for adaptation – the physiological repatterning that allows us to improve - at a certain point in the process induces discomfort. We will look at those processes in more detail below. Given the role of discomfort in adaptation: how do we design systems to help people embrace the discomfort? In HCI, we might say our discipline is designed fundamentally to avoid discomfort. Our focus is often to make engaging with a task or system as effortless as possible. And yet, we also see examples of coping with challenges to progress. Games require skills development to succeed in the game environment where one must embrace the frustration of repeated "deaths" to improve performance. There may be synergies here to the physiological and perceptual discomfort we describe below to draw on.

The motivation for our exploration of discomfort-as-material is two-fold. First, our related work has shown that we seem to have evolved to thrive best when we process discomfort as part of an adaptation [31], so understanding the role of adaptation in our environment and practice is valuable. Second, and what is more the focus of this paper, is that, whether with the use of interactive technology or not, we see people avoid or abandon a practice (like learning statistics) or skill (knee kicking a soccer ball) or state (burning body fat/building lean tissue) that they themselves acknowledge they would like to achieve. We have all likewise heard or experienced that the practice has been abandoned because it is either "boring" or "too hard" or that the person sees "no change" or improvement despite their effort, or they feel they are in pain because of it, and so the practice is abandoned.

Given that discomfort is interlinked with the desired adaption, we ask: what can we learn about discomfort itself and how might we more deliberately embrace and deal with discomfort within our designs? We have elsewhere called the parameters of adjusting practices to support adaptation "tuning"[30]. Here, we look more closely at this particular component of adaptation and how it might be leveraged as a deliberate part of the tuning process. Our rationale for the exploration is that the more we can help each other build knowledge, skills and practice from personal health to social compassion, the more potential there will be for shared joy and social transformation (what has been framed as a call to "#makeNormalBetter"[27]). By designing *with* discomfort awareness, we can help mitigate its negative effects on achieving aspirations.

To be clear, our intent is not to create systems with interfaces that themselves induce discomfort or frustration—such as the interfaces found in notoriously difficult games like QWOP[1]. Rather our intent is to leverage the body of work in HCI to create usable interfaces that support others embracing discomfort for the adaptations they seek. In that sense, discomfort design is both a departure from and an affirmation of usability. It departs from usability by deliberately leaving the difficulty of a practice in place. As we shall see, removing the difficulty would remove the value. It affirms usability by employing HCI principles to make it easier to voluntarily engage in a difficult practice.

Towards this goal, the following sections present the background for this consideration. We then overview the physiology and neurology of adaptation - the processes where discomfort is situationally experienced. This perspective foregrounds three principles: the boundary between discomfort and pain, the need for recovery and the connection between long term value and short term discomfort. Given those principles, we conclude with three issues that designs may consider when incorporating discomfort: preparation for discomfort, recovery assessment and value of the practice. Our intent in sharing this work at this time is to invite the community to explore, test and further refine these parameters towards a generalisable "discomfort design framework" to help us, as a community, design better tools to building in health, wellbeing and performance from individual to infrastructure interactive designs.

## 2 BACKGROUND

Our consideration of discomfort is situated within the context of Inbodied Interaction. Inbodied Interaction [29] fundamentally suggests that by considering the internal processes that take place across all 11 of the body's organ systems, we will be in a place to better *align our designs* with how we function optimally as complex systems of complex systems, operating in complex systems. Discomfort Design was first proposed as a part of this Inbodied Interaction consideration at a Ubiquitous Computing Workshop in 2019 [26] and follow up paper [31]. In that workshop and paper, key consideration for discomfort was how, what we might frame as more global processes of discomfort, seem to create positive effects: being slightly colder at night, going longer during the day without food, being slightly hungry more of the time, dealing with bursts of intense heat and so on, are associated with restorative benefits. How, the workshop participants explored, might this kind of beneficial discomfort be embraced in terms of relations to sustainability - from food consumption to housing [33]? What these questions on this macro level have raised about discomfort include: beyond a sensation of varying degrees of intensity - what *is* discomfort? How is it the same or different than pain? what role does something seemingly negative serve in being fundamentally a part of our progress? And, how can we use these understandings more deliberately in our designs to bring more people into the assumed benefits for which we are designing tools? It is from this context - of attempting to better understand the properties of discomfort as material, that we present the following sections: that discomfort is distinct from pain; that recovery is needed for adaptation and that long term value motivates engagement in short term discomfort.

---

[1] The QWOP game is available at https://www.foddy.net/Athletics.html

## 3   THE ANATOMY OF DISCOMFORT

Discomfort is both a physiologic process and a perception. Discomfort is associated with physiological processes invoked by stimuli and our perception of discomfort is complex. We will consider both process and perception below. We appreciate that the following sections are largely explanatory but necessary as few members of the community have this familiarity with a science of discomfort. Therefore in order to suggest how we can use this as a design material, we offer the following as an overview of properties of that material for interaction design.

### 3.1   Brief overview of The Physiology of Adaptation

In order to understand discomfort, we need to understand something of the fundamental physiological process to which it is attached. That is, adaptation. As we have noted elsewhere, from an inbodied interaction lens, the body is primarily the "site of adaptation" [29]. It manages this adaptation as part of a triple process: adaptation, via metabolism, to maintain homeostasis. In brief, homeostasis refers to the internal environment of the body that must be maintained within strict ranges in order to function. We can track these states via measure like blood pressure, Ph, temperature, fluid levels and so on. Metabolism is the conversion of incoming resources like nutrients into either fuel to enable other metabolic processes like generating new tissues to recycling cells that the body requires to function in response to changing external demands, like learning a new skill, or trying to keep warm in a suddenly cold room. These are responses to demands for adaptation.

   The key takeaway here is that the body is constantly adapting to context. Sitting in a chair, reading this paper, requires a non-stop, always on balancing of internal resources to enable all the demands from maintaining cognitive focus to physical position over any given period of time. Those adaptations will be different depending on previous and ongoing adaptations driven by how we sleep, eat, move, think, engage with others[28]. Within these interactions, the body is constantly responding to the requirements imposed by the environment or context external to the body to maintain the body's internal environment - its homeostasis - in any context. That constant adaptation is managed by metabolism [15].

### 3.2  Discomfort and adaptation

We might ask here: if we are constantly adapting to context, why are we only occasionally discomfortable? We might think of the lack of discomfort this way: the body's metabolism is able to maintain the requirements for more or less familiar physical and cognitive and social demands without requiring significant or out of current range adaptation. Discomfort is generally associated with the experience of pushing the body into a state that challenges its current threshold at which it can maintain homeostasis. That limit will be different for different people, in different contexts.

   Consider someone running as fast as they can to catch a bus pulling away. We may notice that the person slows in about 7-9 secs - and stops before the bus starts to leave; they are huffing and puffing, hands on their knees, though they may only have been moving at a pace not much greater than what is considered walking speed (2-4mph). For that commuter, their discomfort is associated with an unaccustomed effort that has taxed a particular energy system (the phosphocreatine system [11,23]) that can provide a burst of energy for less than 10 seconds, by which point they could physically keep moving, albeit slower (another system, glycolytic, dominates at this time [2]) but, they have told themselves to stop from fear that this unfamiliar pounding in their chest may be a heart attack. Recent work suggests that this discomfort and fatigue that induces quitting [13] is both physical and mental [34]. Their heavy breathing is both rebalancing the gases their body needs for homeostasis, and also panic - a response to a perceived threat rather than actual physiological harm (we often over-breath in recovery from an effort [36]). Often when one is not prepared for or familiar with discomfort associated with a new effort, they will – not unreasonably - interpret it as pain, or boredom, and thus as a signal to stop, rather than as a signal that greater than usual adaptation is actually being invoked right at that moment. Building a skill, building strength, shedding fat, are more obvious cases of adaptation often associated with discomfort: we speak of the amount of discipline and motivation required to keep someone "on task" to put in the "work" required to build new capacity.  We talk about the need to break a plateau in a current skill or state by "getting out of our comfort zone." Discomfort, as a precursor to progress through adaptation, is not optional; the way we experience that discomfort, however, is, and that experience is where we see opportunity for HCI design/research.

### 3.3  Discomfort and Pain

A question that comes up when we discuss discomfort is: what is the difference between discomfort and pain? That is a complex question in no small part because both are complex states. As Melzack in 1999 described the complexity of pain as a "neuromatrix" [19,35], pain is multifactorial. It has multiple contextual components as well as physiologic components. For example, an athlete in the middle of a game may experience a hit or a fall, and feel no pain, get up and keep going.  After the game, the athlete may start to feel pain, and find that they have broken a bone. The brain in these cases stops the signals that would turn on pain perception while it prioritizes other tasks that are perceived to be more critical - like catching a football. That physiologic interaction is informed by context. Pain can be framed as a signal to change. What makes pain challenging in a diagnostic sense is that it doesn't tell us *what* to change once it is felt, because for many reasons, as the well-worn phrase goes, *the site of pain is not the source of pain*.

 To be a little more specific, physical markers of pain include that: it limits performance, affecting range of motion, load or speed at which a movement can be carried out. Thus, in recovery from an injury, pain causes us to adapt movement to work within those constraints as tissues heal. This working around pain can cause neural coordination patterns to change, so that when the area has repaired, rehabilitation work is needed to rebuild the correct nervous tissue patterns to restore function. We often test recovery from pain by checking range of motion, how much load we can move, how quickly we can move it.  We can translate these physical limiters to our social or cognitive lives as well: being at a party with many strangers may cause an unbearable distress that limits social engagement options/success.

 Discomfort has similar components as we will see below, but the difference between sensations that we categorize as either pain or as discomfort is that they often push capability in opposite directions. If an action produces pain, actions that involve that system will be limited by the pain in what they can do - range, speed, load. In fact, the more we carry out a painful action, often the greater the pain will become, with injury increasing, until we stop. With discomfort, doing the same action longer often causes the discomfort to diminish. It may not disappear, but it does not increase or inhibit our speed, load or range of movement. Over frequent exposures, that discomfort becomes less as we adapt to it.

 Some discomfort is also practiced or habituated. Hunger is a great example of this: hunger is a multifactorial experience, but it is strongly associated with a particular hormone, ghrelin, that fires based on habituation - feeding times - rather than around physiological need [1]. As a person extends the period in which they are not eating those signals of discomfort - of the need to eat - change too. In research on fasting, we see leveling out of ghrelin levels that occur through fasting associated with decreased experience of hunger [1]. A challenge many experience - and hence an opportunity for design - is attempting to get through those pangs "cold turkey", that is, without gradual adaptation. That can be challenging and a cause for abandoning a practice as "too hard" when really it is too much discomfort for current adaptive capacity to maintain. It is worth repeating: the discomfort unlike pain does not limit range of motion, speed or load of the activity. But it can cause us to quit.

Like recent research on fatigue, quitting is almost always a mental or cognitive choice within discomfort, rather than metabolic exhaustion [17]. We overview this physiology below.

### 3.4  Process of Discomfort: Recovery from Imposed Load

In adaptations that are usually related to discomfort we can categorize these as a need to adapt to a current state, or to a metabolic threshold. What do we mean by this? For a state adaptation, we can say that we have the current physiologic, metabolic capacity to address, usually as we move from one state to another. Consider changing states from static to active on a cold day. For example, going for a walk on a cold day. We dress such that will likely be a little cold at the start of the walk, coming from a warm indoor space out into a cold outdoor space, but we know that, as we move, we will literally warm up. Our internal state moves from cold to warm. In our walk, we may even begin to feel discomfort from the other end of the temperature scale - needing now to shed clothing, to help maintain a temperature balance relative to our movement state. After some practice, we gain experience both to endure the short period of discomfort and initial cold; and to adjust growing warmth. Our metabolism is active in adjusting our heart rate, circulatory system, air flow to ensure we do "adjust" to the temperature relative to our effort (section 3.5 overviews the role of the Autonomic Nervous System in maintaining these states across contexts).

Where we often see greater discomfort in terms of this physiological (rather than perceptual) experience is where demand approaches physiological thresholds of current capacity. If we use a familiar example of someone building muscle strength, that is associated with what is known as "time under tension" - strength adaptation usually occurs. This strength adaptation affects many processes. One is usually the development of more lean tissue, like muscle fibers, known as hypertrophy. The exact pathways and processes for skeletal muscle hypertrophy is an active area of research, but one thing is pretty well accepted: muscle growth is related to the amount of tension a muscle is able to develop for a particular period. This tension load may be many lifts consecutively until a person can no longer lift the load with the same quality; or it may be a bigger load, for fewer repetitions in that time period again until a person cannot complete the movement with the same quality [14,32].

Based on current models, lifting loads that are beyond a person's norm causes "microdamage" in the body. As a result of this damage - which is often not experienced as pain until the next day - the body responds, as part of the repair process, by creating new tissue - including new muscle fibers. More fibers, more capacity for tension to lift greater load, like adding strands to a rope to make it stronger. It's important to note that this repair process happens during recovery - post damage - and when that recovery is insufficient so is the adaptation. This constant cycle of demand requiring physiological response is the role of metabolism to bring together the necessary materials to support the adaptation to maintain homeostasis.

A similar process occurs for distance running to become experienced as less effortful: one's capacity for bringing air into the lungs, and also for the body's cells to adapt to be able to make use of more air, is also stimulated by carrying out what is known as "threshold training" [16,20,21]. This kind of work is often very demanding. Imagine doing a series of 50-100m sprints, at near your maximal capacity, resting only briefly and doing a set of them again. And then again. The practice is uncomfortable, but this discomfort is part of a signal for the body to build new capacity such that this same amount of effort over time can be invoked with less cost to the body - in other words, so that the process induces less discomfort. That new capacity is built during periods of recovery between training sessions.

This adaptation is critical as a response to demand. Not adapting would mean that the system cannot maintain homeostasis - it cannot repair the challenges to its homeostasis. Research on aging and extending healthy lifespan frames aging in terms of a system's decreased capacity of one's metabolism [10] to meet the repair and recovery challenges.

### 3.5  A Volitional/Perceptual/Inhibitory Aspect of Discomfort and Its Relation to Adaptation

We can frame discomfort as having both *physiologic* and *perceptual* components. Practicing holding one's breath in air hunger drills has physical components of discomfort as the urge to breathe grows. But it also has a different discomfort as we deal with a fear associated with running out of air. We propose to call this type of discomfort *perceptual*. For example, we may always feel similar physical discomfort with air hunger drills, but the perceptual discomfort of fear or panic can subside with experience. Engaging with perceptual discomfort so that it does not unduly inhibit us from exploring a desired practice towards a desired achievement seems a large opportunity space for HCI design research.

It's likely clear from these examples that discomfort can also be experienced from non-physical sources, but whatever the source, discomfort will be experienced physically—as all experience is mediated via the body. A difficult high-stakes ethical challenge, for example, may cause physical responses of associated discomfort such as lack of sleep or poor digestion. These responses may over time challenge our capacity to maintain homeostasis. Challenge homeostasis for sufficient time or intensity, we become ill: a failure of homeostasis to adapt to the discomfortable demand.

Perceptual Discomfort can also be *anticipatory*. These anticipatory and perceptual discomforts do not always inhibit an action or lead to illness. In the physical circumstances outlined in the above sections, like going for a walk on a cold day, we will anticipate discomfort. Based on our familiarity with that discomfortable experience, however, as the expression goes, we may "suck it up" in order to achieve the associated adaptation we seek.

Sucking up generally means that we *choose* to carry out a practice despite the anticipated and/or perceived discomfort. We keep walking in the cold, because (1) experience tells us we will adapt in a short period and (2) experience tells us we can withstand the cold for that long without damage while we warm up. We are balancing perceptions of cost/harm with benefit/adaptation. This process of up sucking discomfort can, however, cause damage from the misguided if popular conceit of "no pain no gain." This is a false corollary. Pain and discomfort as we described above may have similar feeling starting points, but their process and purpose go in near opposite directions. As noted above, pain limits capacity; discomfort, perhaps especially perceptually, seems to be largely a signal for us to pause and reflect on whether this activity is safe to pursue. That is, *how much of a risk to our homeostasis is it?* If we do pursue it, if we do "suck it up" that pursuit will also induce adaptations; our capacities, our thresholds will adapt to these new demands, if done appropriately. Designing discomfort doses that are both safe and effective to support positive adaptation again seems a new opportunity for HCI design. If the discomfort dose is too small, positive adaptation is not likely to take place. If the discomfort dose is too large, the dose becomes pain which reduces capacity.

### 3.6  The Discomfortable Decision: Choosing to Suffer Now

The management of homeostasis is largely run through a part of the nervous system known as the autonomic nervous system (ANS) [8]. The autonomics are those processes that happen without requiring conscious attention. The nervous system itself is divided into two main branches: the central and peripheral [7:1]. The central references the nervous tissue of the spinal cord and the brain. The peripheral references the nerves that connect with every other part of the body. As sensory information like touch, sight, hearing comes from the periphery (PNS) into the central system (CNS), the ANS receives CNS signals from the brain that automatically moderate peripheral processes, like the beating of the heart, the constriction of blood vessels, the depth of breathing. But between the incoming of sensory information from the PNS and the output of autonomic motor responses from the ANS, volitional parts of the brain also come into play. We have a capacity to make decisions based on present perceptions of future value. How far and how accurately we can see into that next moment of discomfort is based on our knowledge, skills and experience – our practice.

In evolutionary biology, this seeing into the future is usually pinned to the actual parts of the brain that support motivation and reward [24]: to be able to see that the discomfort of the present practice towards an adaptation will support a value that is worth more than the discomfort. All of us who have experienced the cost of committing to a project, working to translate the results, having setbacks and recovering from them (including the cycle of publication submission/rejection/revision) is regularly weighing the discomfort of various experiences of a process towards a particular goal. In the CNS we have reward signals that are triggered within particular kinds of efforts, seemingly to help keep us going. We have in our bodies opioid production, not unlike the effects of marijuana, that is triggered during a run, for example, at a certain period and intensity [3] that can help keep us going. Recent work suggests other kinds of euphoric hormonal triggers like dopamine are set off during intellectual pursuits [18]. It's important to note, however, that these reward circuits are not tripped without first going closer to thresholds of capacity - these may be endurance, speed, accuracy, power or for some, thrill [5,25]. In other words, we don't seem to be wired to get the reward without the effort. Our evolution it seems has recognized that discomfort without respite doesn't motivate us to adapt, but reward without effort is likewise life threatening.

**Micro Reward Proximity** Indeed, recent work around motivation, shows that the proximity in time of a reward to the cost of an effort is germane to how much effort a person will put into that effort [9]. For our purposes here, this means that a possible discomfort design vector is to explore how close we can bring an appropriate reward related to the aspiration to the current discomfort effort. For someone who already accepts that effort/discomfort will be a part of becoming stronger in lifting weights at the gym, simply lifting something heavier than the previous day, no matter how hard that experience is, maybe its own micro reward on a larger quest for bigger biceps. It may also result in injury for having an approach to practice more likely to push too hard too fast. Patience can also be discomfortable. How to design to support appropriate but smaller discomfort doses with micro rewards that lead towards larger aspirations?

## 4  DESIGNING WITH DISCOMFORT - PRELIMINARY CONSIDERATIONS.

From a design perspective - what this ancient wiring of ours suggests is that we need to consider how to balance seeing the future, post discomfort and adaptation, with experience of the present. One framing around embracing discomfort is Ericson and Pool's "deliberate practice" [6]– that assumes discomfort. In deliberate practice, every training session includes a deliberate effort to push oneself beyond the current comfort level: moving from repeating tasks that are easy to consistently engaging with tasks that are (currently) difficult. Success of adaptation is measured in progress of performance. They also suggest this work requires a teacher to support feedback necessary for progress, but overall time spent in discomfort (deliberate practice) correlated with degree of expertise built.  This work assumes there will be discomfort but not *what kind or degree* of discomfort, or the cost of perceived discomfort to engagement or persistence. It seems very much of the "suck it up" school of discomfort coping.

We therefore suggest that there are several opportunities for design to explore.  First, improving **Preparation for Discomfort**. This can involve adjusting expectations.  People regularly abandon resistance training activities because in the days *after* lifting - not at the time - they experience intense soreness which can take days to get over. And so, understandably, they abandon the activity. (This abandonment cycle is reflected in the Health Belief Model [22]). And yet, this Delayed Onset Muscle Soreness [12] is a well-known phenomenon.  Correctly understanding that muscle soreness is an expected indicator of positive adaptation may prevent early abandonment of the activity.  We suggest this question of design for Preparation with similar use cases may be valuable to explore at the workshop.

Another opportunity for design exploration is to make explicit the anticipated **Value of the practice,** especially during the short term, with micro rewards that can be assessed relative to the degree of practice.   **Micro rewards** *connect practice with indicators that point toward desired long-term adaptations thus strengthening the connection between present discomfort and future value*.  Many issues need to be resolved in making this connection explicit.  In the workshop, we might consider: Is encouraging more effort now (and possibly more discomfort) supporting the adaptation sought or is it too much effort and becoming pain? Is the micro reward explicitly connected to the desired adaptation strongly enough to trigger motivation? For example, one can cram for an exam if the Value is simply to pass and "suck up" that discomfort where the reward (passing) is closer to the discomfort (results soon after cramming). Cramming may be perceived to induce less discomfort than steady weekly effort/discomfort, especially if there is no apparent periodic micro reward, only the final grade. How might we use such a global, familiar experience as a way to help participants explore/test micro discomfort with microrewards vs such large boluses of discomfort and potentially failure?

A final opportunity is **Recovery assessment**. Adaptation occurs not in the effort but in the recovery. Pacing out practice to interleave effort with recovery (and positive adaptation such as skill building) requires patience and that patience itself can be discomforting. Impatiently, or naively, replacing recovery with additional effort creates pain which limits ability. A misguided "no pain no gain" mentality compounds the problem. Connecting the benefits of recovery to the value of the practice may sustain motivation across periods of inactivity. Several questions arise: How long should the recovery period last? How are the benefits of recovery communicated? How is enthusiasm to act impacted by the need to rest? How is recovery connected to long term value? During the workshop we might consider how to design to support feeling the benefits of recovery as a mitigation for the discomfort of waiting. In endurance training for example, there has been work using HRV as a measure of recovery. How might that be internalized/validated against interoceptive awareness to best anticipate when work/training is next best used? Or similarly, when one is not interesting in training, because the anticipated discomfort of hard effort is off-putting, how might skill deterioration from not embracing that discomfort be reflected? These questions may also be asked around cognitive skills building.

## 5 CONCLUSION & FUTURE WORK

We propose that discomfort as material in HCI can be understood with two particular attributes - like warp and weft - of physiological discomfort and perceptual/anticipatory discomfort. Both strands are related to how the body adapts, particularly how the body adapts to discomfort to create new capacity to be able to maintain homeostasis longer, better in more challenging contexts. Triggering these adaptations involves discomfort rather than pain and includes periods of recovery. Successfully engaging in discomfort and recovery is motivated by connections to perceived long term value.

These principles suggest that new approaches to design which include preparation for discomfort, communicating the future value of discomfort and support for recovery. We suggest that framing discomfort as an explicit design material enables us as researchers and designers in interactive technology to consider more deliberately how discomfort acts to affect engagement with a practice, and thus engagement with designs that support practices. For designs particularly focused on health, wellbeing, creativity - in other words all aspects of human performance - discomfort is therefore a critical, operative factor for such designs. We have sketched out three issues for consideration in discomfort design. These issues range from preparation to recovery.

Both how a design can help one reflect on the balancing of degree of discomfort and the degree of recovery needed, and how to recover from both perceptual and physiological discomfort are uncharted territory. But we suggest, that territory has much to offer to help more people access more knowledge, skills and practices to support their brilliance and resilience. We look forward to developing this exploration of discomfort as material here with the inbodied community, towards more broad uptake in the general HCI research and design communities. Our goal is both to have discomfort embraced as a design material, and also as outcomes from this workshop, to take steps toward a useful and usable general framework for discomfort embracing designs.

## 6 ACKNOWLEDGMENTS

We acknowledge support from EPSRC's GetAMoveON, ReFresh, and Health Resilience Interactive Technology grants and the United States National Science Foundation (NSF) under grant IIS-1406578.

schraefel-jones-inbodied4workshop-2021-chi2021